\documentclass[11pt,twoside]{article}
\usepackage{bm}
\usepackage{latexsym}
\usepackage{dcolumn}
\usepackage{amsmath,amsfonts,amssymb}
\usepackage{graphicx,epsfig}
\usepackage{psfrag}
\usepackage{amsthm}
\usepackage{color}

\usepackage{nccmath}
\usepackage{moresize}
\usepackage{enumerate}
\usepackage[hang,flushmargin]{footmisc}
\usepackage[titletoc,toc]{appendix}
\usepackage{lipsum}
\usepackage{subcaption}
\usepackage{epstopdf} 
\usepackage{hyperref}
\usepackage{rotating}
\usepackage{multirow}
\usepackage{mathtools}

\usepackage{stackengine}
\usepackage{scalerel}

\usepackage{cite}

\newcommand{\be}{\begin{equation}}
\newcommand{\ee}{\end{equation}}
\newcommand{\bea}{\begin{eqnarray}}
\newcommand{\eea}{\end{eqnarray}}
\newcommand{\bse}{\begin{subequations}}
\newcommand{\ese}{\end{subequations}}
\newcommand{\bce}{\begin{center}}
\newcommand{\ece}{\end{center}}
\newcommand{\bfg}{\begin{figure}}
\newcommand{\efg}{\end{figure}}
\newcommand{\bit}{\begin{itemize}}
\newcommand{\eit}{\end{itemize}}
\newcommand{\bed}{\begin{description}}
\newcommand{\eed}{\end{description}}
\newcommand{\ben}{\begin{enumerate}}
\newcommand{\een}{\end{enumerate}}
\newcommand{\nn}{\nonumber}

\newcommand{\pa}{\partial}
\newcommand{\fr}{\frac}
\newcommand{\sq}{\sqrt}
\newcommand{\no}{\noindent}

\def\le {\left}
\def\ri {\right}
\def\a  {\alpha}
\def\b  {\beta}

\def\d  {\delta}

\def\e  {\epsilon}

\def\k  {\kappa}

\def\L  {\Lambda}
\def\m  {\mu}
\def\n  {\nu}

\def\O  {\Omega}

\def\r  {\rho}

\def\s  {\sigma}

\newcommand{\cO}{\mathcal O}

\newcommand{\cR}{\mathcal R}

\newcommand{\nab}{\nabla\!}

\newcommand{\vx}{\vec{\pmb x}}

\newcommand{\rmt}{\r^{(m)}}

\newcommand{\rbt}{\r^{(b)}}

\newcommand{\rct}{\r^{(c)}}

\newcommand{\rBt}{\r^{(B)}}

\newcommand{\UQ}{U_{_Q}}
\newcommand{\rx}{\r_{\!_X}}

\newcommand{\Obt}{\O^{(b)}}

\newcommand{\OLt}{\O^{(\!\L\!)}}

\newcommand{\rmp}{\r^{(m)}_{_0}}

\newcommand{\rbp}{\r^{(b)}_{_0}}
\newcommand{\Obp}{\O^{(b)}_{_0}}
\newcommand{\rBp}{\r^{(B)}_{_0}}
\newcommand{\rcp}{\r^{(c)}_{_0}}
\newcommand{\Ocp}{\O^{(c)}_{_0}}
\newcommand{\OLp}{\O_{_0} \!\!^{\!(\!\L\!)}}
\newcommand{\OBp}{\O^{\mbox{\tiny ({\it B})}}_{_0}}

\newcommand{\Hp}{H_{\text{\tiny 0}}}
\newcommand{\tp}{t_{_0}}

\newcommand*\rfra[2]{{}^{\scriptstyle{#1}}\!\!\diagup_{\!\!\scriptstyle{#2}}}

\newcommand{\bdm}{\begin{displaymath}}
\newcommand{\edm}{\end{displaymath}}

\long\def\symbolfootnote[#1]#2{\begingroup%
\def\thefootnote{\fnsymbol{footnote}}\footnote[#1]{#2}\endgroup}
\begin{document}

\markboth{Saurya Das, Mohit Kumar Sharma and Sourav Sur}
{On the quantum origin of a dark universe}

\title{\LARGE{On the quantum origin of a dark universe
%\\
%A quantum backreacted cosmological dark sector from a Bose-Einstein condensate
}}

\author{Saurya Das\footnote{email: saurya.das@uleth.ca} \\ 
{\normalsize \em Theoretical Physics Group and Quantum Alberta, 
Department of Physics and Astronomy,}\\
{\normalsize \em University of Lethbridge, 4401 University Drive, Lethbridge, Alberta 
T1K 3M4, Canada}\\ \\
Mohit Kumar Sharma\footnote{email: mrmohit254@gmail.com}~ and~
Sourav Sur\footnote{email: sourav.sur@gmail.com} \\ 
{\normalsize \em Department of Physics and Astrophysics}\\
{\normalsize \em University of Delhi, New Delhi - 110007, India}}

\date{}
\maketitle

%%%%%%%%%%%%%%%%%%%%%%%%%%%%%%%%%%%%%%%%%%%%%%%%%%%%%%%

\begin{abstract}

It has been shown beyond reasonable doubt that the majority (about 95\%) of the total energy budget of the universe is given by the dark components, namely Dark Matter and Dark Energy. What constitutes these components remains to be satisfactorily understood however, despite a number of promising candidates. An associated conundrum is that of the coincidence, i.e. the question as to why the Dark Matter and Dark Energy densities are of the same order of magnitude at the present epoch, after evolving over the entire expansion history of the universe. In an attempt to address these, we consider a quantum potential resulting from a quantum corrected Raychaudhuri/Friedmann equation in presence of a cosmic fluid, which is presumed to be a Bose-Einstein condensate (BEC) of ultralight bosons. For a suitable and physically motivated macroscopic ground state wavefunction of the BEC, we show that a unified picture of the cosmic dark sector can indeed emerge, thus resolving the issue of the coincidence. The effective Dark energy component turns out to be a cosmological constant, by virtue of a residual homogeneous term in the quantum potential. Furthermore, comparison with the observational data gives an estimate of the mass of the constituent bosons in the BEC, which is well within the bounds predicted from other considerations.

\end{abstract}

\vspace{10pt}
\no

%\maketitle

%\tableofcontents

%%%%%%%%%%%%%%%%%%%%%%%%%%%%%%%%%%%%%%%%%%%%%%%%%%%%%%%

\section{Introduction}

A unified picture of the cosmic Dark Matter (DM) and Dark Energy (DE) has been one of the 
key aspirations of modern researches within the standard paradigm of spatially flat 
Friedmann-Robertson-Walker (FRW) cosmology
\cite{CST-rev,AT-book,wols-ed,MCGM-ed,BCNO-rev}. 
While there have been innumerable attempts to emulate the nature and properties of DM and DE, 
a general consensus on the specific structure of the evolving dark sector remains elusive till 
date. Although candidates abound, e.g. WIMPs and axions for the DM
\cite{wimp1,axion1,axion2,axion3,axion4},
and a plethora of others for the DE (such as scalar fields, aerodynamic fluids and modified 
gravitational artefacts)
\cite{CDS-quin,carroll-quin,CLW-quin,ZWS-quin,tsuj-quin,PMS-kess1,PMS-kess2,MCLT-kess,
scher-kess,SD-kess,PT-dil,BJP-tach,CL-tach,MM-tach,BBS-chap,XLW-chap,NO-mg1,NO-mg2,
SF-mg,clif-mg},
the emergent models have had their own pros and cons. One conundrum, of course, is the 
`{\em coincidence}', i.e. the vast amount of fine-tuning required to set up the initial 
conditions for the same order-of-magnitude contributions of DM and DE at the present epoch, 
if they purportedly stem out of different fundamental sources. This is indeed perplexing, 
given the strong observational support for the DE density to be closely akin to a cosmological 
constant $\L$, whereas almost the entire bulk of the DM to be `cold' and non-relativistic, 
with energy density $\rct (t) \sim a^{-3} (t)$ (where $a(t)$ is the cosmological scale factor, 
and $t$ denotes the cosmic time coordinate).

One way to resolve the problem of coincidence (or, at least ameliorate it to a certain degree) 
is to make allowance for interaction(s) between the DE and DM components
\cite{amend-DEM,CPR-DEM,FP-DEM,CW-DEM,BBM-DEM,BBPP-DEM,LZ-DEM,GNP_DEM}.
However, these 
components being so different in character, the physical realization of such interactions is 
often very difficult, unless they appear naturally, for e.g. via conformal transformations in 
the scalar-tensor formulations of gravitational theories
\cite{fuj-st,frni-st,ENO-st,FTBM-st,BGP-st,SB-MST,BS-MST,SS-MST}.
A more robust approach therefore is to bring a {\em unified} dark sector in the reckoning, 
i.e. to consider a scenario in which the DE, the DM, and possibly certain interaction(s) 
thereof, all having the same fundamental origin
\cite{CM-mm,CMV-mm,SVM-mm,CAS-MMT,SAC-MMT}.
Of course, in that way the DE and the DM would lose their individuality, and would merely 
be the structural artefacts of the source, which thereby would become the all-important 
entity that shapes up the evolution profile of the cosmological dark sector.

In this paper, we resort to such a unified dark universe picture, by adopting an earlier 
proposal
\cite{sd,db1,db2,db3},
albeit in a much broader sense. Specifically, we endeavor to deal with the plausible 
perception of the dark sector being described in entirety by a {\em Bose-Einstein condensate} 
(BEC) of light bosons extending across cosmological length scales. The density of the BEC, 
together with its {\it quantum potential}, manifest as the effective DE and DM densities. It 
is worth noting that in principle, every wavefunction comes associated with it a quantum 
potential, which effectively gets added to any starting classical potential, in determining 
the dynamics of the system --- a fact that is often overlooked. In the scenario we focus on, 
the macroscopic BEC provides the requisite wavefunction.

\section{BEC Cosmological Formulation in the standard setup} \label{sec:BEC-Cosm}

Let us begin with the general form of the cosmological equations in the standard 
spatially flat FRW framework:
\bea 
&& H^2(t) \equiv \le[\fr{\dot a(t)}{a(t)}\ri]^2 =\, \fr{\k^2} 3 \, \r\,(t) \,\,, 
\label{cosm-eq1}\\
&& {\dot H}(t) +\, H^2(t) \equiv\, \fr{\ddot a(t)}{a(t)} =\, -\, \fr{\k^2} 6 
\Big[\r\,(t) \,+\, 3 p\,(t) \Big] 
\,, \label{cosm-eq2}
\eea
where $\k^2 = 8\pi G$ is the gravitational coupling constant, and $a(t)$ is the 
FRW scale factor, normalized to unity at the present epoch, $t=t_0$, i.e. $a(\tp) 
= 1$. The overhead dot $\{\cdot\}$ denotes $\pa/\pa t$, and $\r\,(t)$ and $p\,(t)$ 
are respectively the conserved total energy density and pressure of the system. 
The latter is presumably a collection of fluids and(or) fields, the individual 
energy densities and pressures of which may not necessarily be conserved. 

Eqs.\,(\ref{cosm-eq1}) and (\ref{cosm-eq2}) are of course classical equations. 
However, if a constituent entity is inherently quantum mechanical, and described 
by a wave-function of the form 
\be \label{WF}
\Psi (\vx,t) =\, \cR (\vx,t) \, e^{i \, S (\vx,t)} \,, \qquad \mbox{with} \quad
\cR, S \in \mathbb{R}\,\,,
\ee
then as shown in
\cite{sd,db1,db2},
an additional term arises due to the quantum correction of the field equations, 
even at the background cosmological level. Specifically, the geodesic flow equation, 
or the Raychaudhuri equation, (\ref{cosm-eq2}) gets modified by a {\em quantum 
potential} given by 
\cite{sd}
\be \label{QP}
\UQ =\, \fr{\hbar^2}{m^2} \, h^{\m\n} \, \nab_\m \nab_\n \le(\fr{\square \cR} 
\cR\ri) \,,
\ee 
with $m$ being the total mass parameter of the quantum constituent, and $\,h_{\m\n}$ 
denoting the (induced) metric of the three-dimensional spatial hypersurfaces of 
constant cosmic time. One can, as usual, express $\,h_{\m\n} = g_{\m\n} + u_\m 
u_\n\,$, where $\, u_\m = (\hbar/m) \,\pa_\m S \,$ is the world velocity of the 
evolving (quantum) cosmic `fluid', which can appropriately be taken to be a BEC of 
rest mass $m$ and described by a macroscopic wave-function of the form (\ref{WF}), 
provided the value of $m$ remains within certain limit 
\cite{db1,db2}.

Indeed, for a suitable choice of the amplitude $\cR$, the BEC probability density 
$\rBt \sim |\cR|^2$ is shown to mimic the energy density of a classical dust-like 
matter, presumably the cold dark matter (CDM), whereas the potential $\UQ$ is 
positive (repulsive) and even a constant, that acts as a cosmological constant 
dark energy (DE) component $\L$, albeit of a quantum origin
\cite{db2,db3}.
However, from a rather general perspective, such a CDM mimicry of $\rBt$ may only 
be partial, since the potential $\UQ$ may evolve dynamically and contribute to the 
CDM as well. That is to say, $\rBt$ or $\UQ$ may not individually be responsible, 
in general, for shaping up the course of evolution of only the CDM or only the DE. 
Instead, their combined effect may be looked upon as what leads to some specific 
evolution profile of the entire cosmological dark sector. Strikingly however, 
one may still reckon the constancy of the effective DE component, and that the
CDM effectively results from a close interplay of not only the evolving $\rBt$ 
and $\UQ$, but also the density of the visible (baryonic) matter content of the 
universe. While demonstrating all these in this paper, we shall endeavor to make
it clear that the formalism developed herein differs substantially from the
plethora of proposals encompassing the ways to treat the BEC as a plausible CDM 
candidate in the literature
\cite{sahni,hu,Lopez,Bohua,sudarshan1,sudarshan2,khlo1,khlo2,morikawa,moffat,wang,
boehmer,sikivie,harko,chavanis,dvali,houri,kain,suarez,ebadi,laszlo1,laszlo2,bettoni,
gielen,schive,davidson,casadio1,casadio2}. 
%. 
In particular, it is worth noting here that, to the best of our knowledge, 
neither the quantum effects of the BEC on the background level cosmic 
evolution, nor the possibility of a BEC accounting for the whole dark sector, 
has been explored with rigour. 

Let us first recollect an important result derived in 
\cite{db1,db2}:
{\em for a copious supply of bosons of mass $\lesssim 6\,$eV, the corresponding 
critical temperature $T_c$ exceeds the ambient temperature of the universe at 
all epochs}. To be more precise, the critical temperature of an ideal gas of 
ultralight bosons of mass $m$, below which they will drop to their lowest energy 
state and form a BEC is found in 
\cite{db1}, 
following the procedure outlined in 
\cite{brack,grether,fujita},
\be
T_c (t) \simeq\, 4.9 \, m^{-1/3} \, a^{-1}(t) \quad[\mbox{in $K$, for $m$ in
eV}] \,\,,
\label{ct1}
\ee
under the assumption that the BEC density $\rBt$ accounts for the DM density
in entirety. Consequently, for $m \lesssim 6\,$eV, one has $\, T(a)< T_c(a) \, 
\forall \,a \,$, where $\, T(a) = 2.7 \, a^{-1} \,$ is the ambient temperature 
of the universe. This therefore points to the formation of a BEC of the bosons 
in the early phase of evolution of the universe, once the BEC density is taken
to be of the order of the critical density of the universe at the present 
epoch
\footnote{Note however the difference between the expression (\ref{ct1}) 
and that derived in \cite{haber}. This is because the former is obtained upon 
considering the BEC of neutral bosons, which are their own anti-particles, with 
no charge or other hidden quantum numbers and a vanishing chemical potential 
(e.g. gravitons or axions). In contrast, the authors in \cite{haber} have 
considered charged bosons with a non-vanishing chemical potential, distinct 
particles and anti-particles and a net BEC charge.}.

Now, the observed isotropy of the universe (at large scales) makes the stringent 
demand that the wave-function of this embedded BEC has to be spherically symmetric 
after all. Henceforth, from an analogy commensurable with a limiting Newtonian 
cosmological approximation (see refs.\cite{db2,db3}), we shall assume such a 
wave-function to be of the form: 
\be \label{WF1}
\Psi (r,t) =\, R(t) \, e^{-r^2(t) /\s^2} \, e^{- i E_0 t/\hbar} \,, 
%\label{psi1}
\ee
where $\s$ and $E_0$ are real and positive-valued constants, and
\be \label{rad}
r(t) = \sq{h_{ij} (t) \, x^i x^j} \,=\, x \, a(t) \,, 
\ee
with $\, x = \sq{\d_{ij} x^i x^j}\,$ denoting the comoving radial coordinate.
Eq.\,(\ref{WF1}) essentially means taking the quantum amplitude $\cR (r,t)$ in 
Eq.\,(\ref{WF}) to be a Gaussian (of spread $\s$), modulo a time-dependent 
factor $R(t)$. On the other hand, the phase is considered to be purely a 
function of time, viz. $S (t) = E_0 t/\hbar$, with the parameter $E_0$ having 
the usual interpretation of the ground state energy of the BEC. Of course, in 
an effective cosmic fluid description, one requires $E_0 = m\,$ --- the rest 
mass energy of the fluid, in order that the world velocity $u_\m$ satisfies 
the normalization condition $\, u_\m u^\m = -1$. 

Note also that the above equation (\ref{rad}) is just the standard relationship
between the radial coordinate distances in the proper frame and the comoving
frame, however with the induced metric being identified with that of the 
three-dimensional spatial hypersurfaces of the FRW bulk space-time, i.e. $h_{ij} 
\equiv a^2(t) \d_{ij}$. This is justified from the point of view that we are
simply following the usual course one takes while studying the cosmic evolution 
driven by an intrinsically inhomogeneous source (in this present context, the 
BEC). To be more specific, the local inhomogeneities being usually treated as 
small perturbations over a homogeneous FRW background, it is natural for us to 
make such an identification, while limiting our attention to the background 
level BEC cosmology in this paper\footnote{A rather rigorous study of the 
consequences of the BEC and the quantum potential at the level of the linear 
perturbations is being carried out in a subsequent work \cite{sdmsss-QB}.}. 
The foremost task 
that remains then is to solve explicitly the background cosmological equations 
for the scale factor $a(t)$ (or at least, the expansion rate $H(a)$), which 
would nonetheless require only the homogeneous (or the purely time-dependent) 
parts of the BEC density $\rBt$ and the quantum potential $\UQ$. Henceforth, we 
shall make no allusion to the $x$-dependence of these quantities, or in other
words, denote by $\rBt$ and $\UQ$ their homogeneous parts only. The former is 
of course given simply by
\be \label{rhoBt} 
\rBt (t) =\, | R(t) |^2 \,\,,
\ee
whereas the latter, derived from the above expression (\ref{QP}), turns out to 
be
\be \label{QPt}
\UQ (t) =\, \fr{3 \hbar^2}{m^2} \le[H(t) \, \dot{Y}(t) \,+\, \fr 4 {\s^2}
\le\{F(t) +\, \fr 2 {\s^2}\ri\}\ri] \,,
\ee
with the functions $Y(t)$ and $F(t)$ defined as follows:
\bea
Y(t) &:=& \fr{\ddot{R}(t)}{R(t)} +\, \fr{3 H(t) \,\dot{R}(t)}{R(t)} \,\,, 
\label{Wt} \\
F(t) &:=& \dot{H}(t) +\, 5 H^2 (t) +\, \fr{2 H(t) \,\dot{R}(t)}{R(t)} \,\,.
\label{Ft}
\eea
Clearly, the emphasis is therefore on the function $R(t)$ in the BEC wave amplitude, a suitable form of which leads to a viable cosmological scenario with 
a unified dark sector, as demonstrated in what follows.

\section{BEC Cosmological Evolution and the Unified Dark Sector} \label{sec:BEC-DS}

For the BEC to emulate an effective CDM evolution (either totally or partially),
the corresponding (homogeneous) density $\rBt (t)$ requires to fall off with time, 
i.e. with the expansion of the universe. In particular, with the wave amplitude 
having the functional dependence 
\be \label{Rt}
R(t) =\, R_{_0} \, a^{- 3/2} (t) \,, \quad \mbox{where} \quad R_{_0} \equiv R(\tp) 
: \, \mbox{constant}\,,
\ee 
we have the BEC evolving precisely as a dust-like matter, viz.
\be \label{rhoB} 
\rBt (t) =\, \rBp \, a^{-3} (t) \,, \quad \mbox{with} \quad \rBp \equiv \rBt(\tp) 
: \, \mbox{constant}\,,
\ee
at the background cosmological level. The (homogeneous) quantum potential $\UQ (t)$ 
is then given by Eq.\,(\ref{QPt}), albeit with the constituent functions reducing to
\bea
Y(t) &=& -\, \fr 3 2 \le[\dot{H}(t) +\, \fr{3 H^2(t)} 2\ri]\,, 
\label{W1} \\
F(t) &=& \dot{H}(t) +\, 2 H^2 (t) \,\,.
\label{F1}
\eea
Consequently, the Friedmann equations (\ref{cosm-eq1}) and (\ref{cosm-eq2}) yield
\bea 
\r\,(t) &=& \fr 2 {\k^2} \Big[2 Y(t) \,+\, 3 F(t)\Big] \,, \label{rhotot} \\
p\,(t) &=& \fr 4 {3\k^2} \, Y(t) \,\,, \label{prtot}
\eea
so that we can re-write Eq.\,(\ref{QPt}) as
\be \label{QP1}
\UQ(t) =\, \fr{24 \hbar^2}{m^2 \s^4} \le[1 \,+\, \fr{\k^2 \s^2}{12} \le\{\Big[1 
\,+\, \fr{3 \k^2 \s^2} 8 \, a(t) \, p'(t)\Big] \r\,(t) \,-\, 3 p\,(t)\ri\}\ri] \,,
\ee
where the prime $\{'\}$ denotes derivative with respect to the scale factor $a(t)$. 
Eq.\,(\ref{QP1}) is a convenient form of the quantum potential, which 
explicitly shows how it depends on the physical variables, viz. the total (or the
`critical') energy density $\r\,(t)$ and the total pressure $p\,(t)$.

Now, since we are primarily interested in the late-time expansion history of the
universe, it suffices to consider the visible matter to be baryonic dust, having 
an energy density
\be \label{rhob} 
\rbt (t) =\, \rbp \, a^{-3} (t) \,, \quad \mbox{with} \quad \rbp \equiv \rbt(\tp) 
: \, \mbox{constant}\,.
\ee
More specifically, it is reasonable to resort to the following decomposition of the 
total energy density:
\be \label{rhotot1}
\r \,(t) =\, \rbt (t) +\, \rct (t) +\, \rx (t) \,\,,
\ee
where we denote the total effective CDM density by
\be \label{rhoc} 
\rct (t) =\, \rcp \, a^{-3} (t) \,, \quad \mbox{with} \quad \rcp \equiv \rct(\tp) 
: \, \mbox{constant}\,,
\ee
and the total effective (and possibly dynamical) DE density by $\rx (t)$. Note that
$\rct (t)$ may not necessarily be equal to $\rBt (t)$, i.e. the BEC density may not
be the total CDM density, as some contribution may come from the the quantum potential
$\UQ$ as well. Nevertheless, our main interest is in the evolution of the total
dark sector density, $\rct (t) + \rx (t)$, while the corresponding pressure is that
due to the DE component only. In fact, this pressure is the lone contribution to 
the total pressure $p\,(t)$, as the visible matter is dust-like. 

On the other hand, the quantum corrected Raychaudhuri equation for the system 
constituted by only the (visible) baryonic dust and the BEC (which also emulates
a dust-like fluid) is given by
\be \label{QRE1}
\fr{\ddot a(t)}{a(t)} =\, -\, \fr{\k^2} 6 \Big[\rbt(t) +\, \rBt(t)\Big] +\, 
\fr 1 3 \, \UQ (t) \,\,.
\ee
Comparing this with the Friedmann equation (\ref{cosm-eq2}), and using 
Eqs.\,(\ref{rhoB}) and (\ref{rhob}), we find 
\be \label{rhotot2}
\r\,(t) =\, \fr{\rbp +\, \rBp}{a^3(t)} -\, \fr 2 {\k^2} \, \UQ (t) -\, 3 p\,(t) \,\,.
\ee
Hence, the above equation (\ref{QP1}) can be reduced to that of the quantum potential
expressed (for convenience) as a function of the scale factor $a(t)$: 
\be \label{QP2}
\UQ(a) =\, \fr{6 \,\a^2}{1 +\, \a^2 \s^2 \,f(a)} \le[1 \,+\, \fr{\k^2 \s^2}{12} 
\le\{\fr{\big[\rbp +\, \rBp\big] f(a)}{a^3} \,-\, 3 \big[1 + f(a)\big] p\,(a)\ri\}\ri] \,,
\ee
where we have used the following notation and definition:
\be \label{fa}
\a \,\equiv\, \fr{2 \hbar}{m \s^2} \,\,, \qquad
f(a) :=\, 1 +\, \fr{3 \k^2 \s^2 \, a \,p'(a)} 8 \,\,.
\ee
%
%\bea \label{QP2}
%\UQ(a) &=& \fr{24 \hbar^2}{m^2 \s^4} \le[1 \,+\, \fr{4 \hbar^2}{m^2 \s^2} 
%\le\{1 \,+\, \fr{3 \k^2 \s^2} 8 \, a \, p'(a)\ri\}\ri]^{-1} \times \nn \\ 
%&&~~~ \le[1 \,-\, \fr{\k^2 \s^2} 2 \le\{1 \,+\, \fr{3 \k^2 \s^2}{16} \, a \, 
%p'(a)\ri\} 
%+\, \fr{\k^2 \s^2}{12} \le\{1 \,+\, \fr{3 \k^2 \s^2} 8 \, a \, p'(a)\ri\}
%\le(\fr{\rbp +\, \rBp}{a^3}\ri)\ri] \,.
%\eea
%
Again, from Eq.\,(\ref{rhotot2}) and the Friedmann equations (\ref{cosm-eq1}) and 
(\ref{cosm-eq2}) (or more specifically, the conservation relation $\, \r'(a) = - 
(3/a) \big[\r\,(a) +\, p\,(a)\big] \,$) it follows that the quantum potential can 
also be expressed in an integral form as
\be \label{QP3}
\UQ(a) =\, \fr{\k^2}{2 a^3} \le[\rBp \,-\, 3 \int\! da \cdot a \le\{a^2 p\,(a)\ri\}'\ri] \,.
\ee
So, for a given form of $p\,(a)$ we can in principle determine $\UQ(a)$, and hence the 
the total density $\r\,(a)$, or equivalently the Hubble rate $H(a)$ that describes the 
cosmic expansion history (and its future extrapolations). 

Let us now consider a rather simplified BEC cosmological setup with the total pressure 
\be \label{LCDM-pr}
p \,= \, - \, \L \,=\, \mbox{constant} \,.
\ee
That this is indeed tenable and the corresponding system of equations, given explicitly
by the above set (\ref{rhotot2}) -- (\ref{QP3}), do admit an exact solution describing 
an effective $\L$CDM evolution had already been demonstrated in a few earlier works 
involving one of us (SD) \cite{db1,db2,db3}. However, certain inherently intriguing 
aspects of such a solution remain to be explored, particularly from the point of view 
of asserting stringent parametric limits pertaining to the physical realization of a 
unified cosmological dark sector. We endeavor to do so in this paper, by first noting 
that Eq.\,(\ref{LCDM-pr}) immediately implies Eq.\,(\ref{QP3}) reducing to
\be \label{LCDM-QP}
\UQ(a) =\, \k^2 \L \le(1 +\, \fr \b {a^3}\ri) \,,
\ee
where $\b$ is a (dimensionless) integration constant. Comparing Eqs.\,(\ref{QP2}) and
(\ref{LCDM-QP}) we identify
\be \label{Lambda-beta}
\L =\, \fr{6 \, \a^2}{\k^2 \le(1 -\, 2 \a^2 \s^2\ri)} \,\,, \qquad
\b =\, \fr{\a^2 \s^2 \le[\rbp +\, \rBp\ri]}{2 \le(1 +\, \a^2 \s^2\ri) \L} \,\,,
\ee
since, by Eq.\,(\ref{fa}), we have $f = 1$ for constant $p$. It then follows from
Eq.\,(\ref{rhotot2}) that the total energy density of the universe is given by
\be \label{LCDM-rho}
\r\,(a) \,=\, \fr{\rmp}{a^3} \,+\, \L \,\,,
\ee
i.e. apart from a cosmological constant $\L$, the universe only has a dust-like matter content 
of energy density $\, \rmt(a) =\, \rmp a^{-3} \,$, whose value at the present epoch 
($t = \tp$ or, $a = 1$) can easily be identified as
\be \label{LCDM-rm0}
\rmp \,= \le(1 -\, \e\ri) \le(\rbp +\, \rBp\ri) \,, 
\ee
where the dimensionless quantity
\be \label{eps}
\e \,=\, \fr{\k^2 \s^2 \L}{3 \le(2 +\, \k^2 \s^2 \L\ri)} 
\,=\, \fr{\a^2 \s^2}{1 \,+\, \a^2 \s^2} \,\,,
\ee 
is positive definite and of particular importance in the context of our entire 
analysis in this paper. Specifically, the above solution (\ref{LCDM-rho}) is the 
same as the one that describes the standard $\L$CDM evolution of the universe 
composed of a dust-like baryonic matter plus the CDM, and dark energy in the form 
of the constant $\L$. However, the key point to note here is that the effective CDM 
density at the present epoch, viz. 
\be \label{LCDM-rc0}
\rcp \,=\, \rmp -\, \rbp \,= \le(1 -\, \e\ri) \rBp -\, \e \, \rbp \,\,,
\ee
is not entirely given by the present-day value $\rBp$ of the BEC density, but is 
actually {\em smaller} than the latter. Interestingly, the reduction is not only 
given by an amount $\e \rBp$, but also by an amount $\e \rbp$, where $\rbp$ is the 
present-day value of the baryonic energy density. Now, recalling that the baryons
have primarily been supposed to constitute only the visible matter content of the 
universe, their role in inflicting a reduction in the effective CDM density (from 
that of the BEC) is nonetheless quite striking. Such a reduction can in fact be 
looked upon as the consequence of the baryons backreacting on the geometrical 
structure of the space-time, because of the quantum correction to the Raychaudhuri 
equation which leads to the quantum potential $\UQ$. Therefore, notwithstanding 
the classical dust-like characterization of the baryonic matter, one can infer 
that the effect of the baryons on the CDM is inherently quantum mechanical. The 
extent of such an effect is crucially determined by the parameter $\e$, given by
Eq.\,(\ref{eps}), which we shall henceforth refer to as the {\em quantum 
backreaction} (QB) parameter. Let us express this parameter in a rather 
convenient form as
\be \label{eps1}
\e \,=\, \fr{\Hp^2 \,\s^2 \,\OLp}{2 \,+\, 3 \Hp^2 \,\s^2 \,\OLp} \,\,, 
\ee 
where $\OLp$ denotes the value of the $\L$-density parameter $\, \OLt (t) = 
\L/\r\,(t) \,$ at the present epoch ($t = \tp$), and $\, \Hp \equiv H (\tp) \,$ 
is the Hubble's constant. It is then straightforward to obtain the following 
stringent bounds being imposed on the extent of the QB from entirely physical 
considerations:
\bit 
\item First of all, the total matter density of the universe has to be positive 
definite. Therefore, apart from being positive-valued by definition, the parameter 
$\e < 1$, so that by Eq.\,(\ref{LCDM-rm0}), $\rmp > 0$ (of course, under the 
presumption that $\rbp > 0$ and $\rBp > 0$, which in turn imply $\rcp > 0$ as 
well as $\rbp < \rmp$).
\item Now, from Eq.\,(\ref{eps1}), the second equality in Eq.\,(\ref{eps}) and 
the definition of the constant $\a$ given by the first equation in (\ref{fa}), we 
have the BEC mass parameter expressed as
\be \label{BEC-mass}
m \,\simeq\, \fr{\hbar \, \Hp} \e \sq{2 \le(1 -\, \e\ri) \le(1 -\, 3 \e\ri) \OLp} \,\,,
\ee 
whose real-valuedness (i.e. $m^2 > 0$) makes evident the more restrictive condition 
$\, \e <\, \rfra 1 3\,$. 
\item Moreover, the overall credibility of a BEC cosmological formulation demands that 
the energy density due to the BEC should not exceed the total matter density. 
Referring therefore to the present epoch, we should have $\rBp < \rmp$, so that by 
Eqs.\,(\ref{LCDM-rho}) and (\ref{LCDM-rm0}), 
\be \label{eps-bound}
\e \,<\, \fr{\Obp}{1 -\, \OLp +\, \Obp} \,\,,
\ee
where $\Obp$ is the present-day value of the baryon density parameter $\, \Obt(t) = 
\rbt(t)/\r(t) \,$. Statistical estimations using recent observational data from various
probes fairly conform to the best fit parametric values for $\L$CDM to be $\OLp \simeq
0.7$ and $\Obp \simeq 0.05$, whence by Eq.\,(\ref{eps-bound}), $\e \lesssim \rfra 1 7$. 
For instance, the widely accepted Planck 2018 results for the Cosmic Microwave Background 
(CMB) anisotropy observations (power spectrum TT,TE,EE+lowE), in combination with Lensing 
and Baryon Accoustic Oscillations (BAO), show the estimates (upto the $68\%$ confidence 
limits): 
\bea \label{Pl18-comb-est}
&& \OLp \,=\, 0.6889 \,\pm\, 0.0056 \,\,, \qquad \mbox{and} \nn\\
&& \Obp h^2 \,=\,0.02242 \,\pm\, 0.00014 \,\,, \quad
\Ocp h^2 \,=\,0.11933 \,\pm\, 0.00091 \,\,,
\eea 
with $\, 100 \,h \equiv \Hp \,$ [Km s$^{-1}$ Mpc$^{-1}] =\, 67.66 \,\pm\, 0.42 \,$.
Therefore, plugging in the best fits of $\OLp$ and $\Obp$ in Eq.\,(\ref{eps-bound}) we 
get $\, \e < 0.136 \,$, which is an even tighter constraint (compared to $\e < \rfra 1 3$ 
found earlier).
\eit
The most interesting outcome of this smallness of $\e$ is of course the enhancement of the 
BEC mass $m$ from its `Hubble value' ($m_{_H} \simeq 10^{-32}$~eV)\footnote{The Hubble value 
is nothing but the one obtained by assuming the BEC constituents to be the gravitons, whence 
taking the Gaussian spread $\s \simeq \Hp^{-1}$ one gets $m \equiv m_{_H} \simeq 2\sq{2} \Hp 
\simeq10^{-32}$~eV, as $\Hp \simeq 10^{-42}$~GeV in the units $c = 1 = \hbar$ (see ref.
\cite{db2}).
}, since by Eq.\,(\ref{BEC-mass}), $m$ varies approximately as $\e^{-1}$. 
%
%%%%%%%%%%%%%%%%%%%%%%%% figure %%%%%%%%%%%%%%%%%%%%%%%%
\begin{figure}[htb]
\centering
   \includegraphics[scale=0.65]{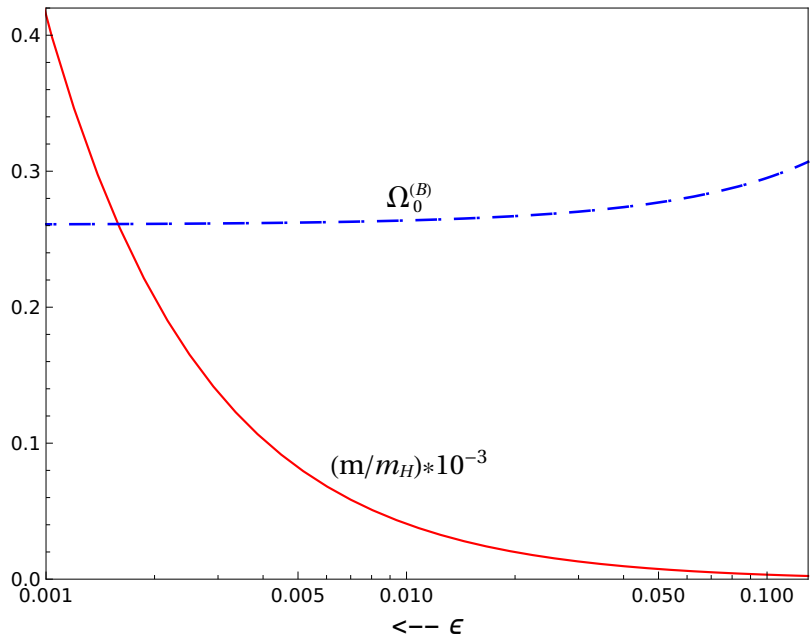}
   \caption{\footnotesize Enhancement of the BEC mass parameter $m$, relative to its Hubble
   value $m_{_H}$ (scaled by a factor of $10^3$), and the variation of the BEC density 
   parameter at the present epoch, $\OBp$, with the decrease of the quantum backreaction 
   parameter $\e$ from about $0.1$ to $0.001$. The $\e$-axes ticks are in the log scale.}
\label{BEC-fig}
\end{figure}
%%%%%%%%%%%%%%%%%%% figure ends %%%%%%%%%%%%%%%%%%%%%%%%%
%
Hence, the smaller 
the value of $\e$, the greater would be the order-of-magnitude enhancement of $m$, which is 
in fact desirable from the point of view of preventing small-scale structure formation of 
the CDM because of the uncertainty principle 
\cite{hu}.
Such a relative mass enhancement with decreasing $\e$ (from its limiting value of about $0.1$ 
found above) is shown in Fig.\,\ref{BEC-fig}, alongwith the $\e$-variation of the BEC density 
parameter at the present epoch, $\OBp \equiv [\rBt/\r]\big\vert_{t = \tp}$, obtained from 
Eq.\,(\ref{LCDM-rc0}). In all the
requisite calculations for these plots, we have used the best fit values of the 
Planck 2018 TT,TE,EE+lowE+Lensing+BAO parametric estimations for $\L$CDM quoted in the
Eq.\,(\ref{Pl18-comb-est}) above.

\section{Conclusion and Outlook}

To summarize, in this paper we have studied a macroscopic BEC stretching across cosmological
length scales as a potential CDM candidate. While there have been a number of attempts in 
this direction earlier, what is new in our work is the observation that the quantum potential
of this BEC gives rise to the effect of an observed DE in the form of a cosmological constant
$\L$. In fact, more intriguingly, we not only perceive a coveted unification of the CDM and 
the DE, as they stem out of the same source, but also that the visible baryonic matter 
content of the universe crucially backreacts on the metric to determine ultimately the
evolutionary profile of the cosmic dark sector. Note that the positivity and the smallness 
of $\L$ here can be attributed to the fact that the quantum potential for the given
wavefunction being treated as a classical potential (or part thereof) in a gravitational
formulation is itself positive. On the other hand, although simple physical considerations,
such as the positive-definiteness of the total energy density of the universe and the 
squared BEC mass ($m^2$), can assert the extent of the quantum backreaction (QB), the most
stringent bound on the corresponding parameter $\e$ comes from the legitimate demand that 
the BEC density should not exceed the total density of the universe at the present epoch. 
As a by-product, albeit of considerable importance, the upper bound on $\e$ enables us to
determine the lower bound on $m$. In particular, it is well-known that the preferred boson 
mass in the BEC, that can prevent the formation of small-scale structures in CDM via the
uncertainty principle
\cite{hu},
is $m \sim 10^{-22}~eV$. While the above stringent bound is found to be $\, \e \lesssim
\cO(10^{-1})\,$, it is ensured thereby that the lower bound on $m$ is well within the 
allowed mass range for the BEC (i.e. between $10^{-22}$ eV and $10^{-32}$ eV, the Hubble
value). Of course, on the whole, it must be pointed out here that it is the macroscopic
wavefunction of the BEC, and the specific form of that wavefunction considered here, 
which have made it essential for all the above to work.
%
%The slight excess of 
%$\rho_{\Lambda}$ over 
%$\rho_{DM}$ 
%in reality and the resulting accelerating universe
%can be attributed to the continuous dilution of DM, 
%$\omega$ not being strictly a constant, that not 100\% of the bosons are in the ground state
%and that we have ignored self-interaction of the bosons. 
%We assumed a slowly varying $\rho_{crit}$, which is guaranteed as soon as the quantum potential is generated, since 
%$\rho=\rho_{DM} + \rho_{DE}$ and although the former decays as $1/a^3$, the latter remains constant in time.
%That said, the assumption of the 
%wavefunction (\ref{psi1}) is bound to  break down at late times. 
%
%This means that there will be a change, and in particular a decrease in the value of $\Lambda$ over time, possibly in discrete jumps. This and the consequent slight increase in the estimated age of the universe should have observable consequences. 
%{Our model is of course consistent with 
%current cosmological observations stretching to 
%a redshift $z\approx 1,100$. 
%No change of
%$\Lambda$ has been observed for this range, which is consistent with our $\omega^2 \approx $ constant assumption over times much smaller compared to the inverse Hubble scale, corresponding to redshift in excess
%of $z\approx 10^{25}$ during the inflationary epoch.
%We have also ignored the backreaction of the BEC on the metric, which should remain valid for the small boson mass required for the BEC to form in the first place. This allows one to study the Klein-Gordon equation, without it being coupled to the Poisson equation with the BEC as its source.
%
%
Nevertheless, we have not contemplated on any deviation from the spatial flatness 
assumption, and have accepted the latter as an observational fact, while determining 
the bound on the QB parameter $\e$, or that on the BEC mass $m$. Moreover, we have not 
looked into the detailed aspects of the BEC, specifically, what its constituent bosons 
would be. Such an exploration is certainly worthy, from the point of view of deriving 
the density profiles of the CDM and comparing the same with the galaxy rotation data. 
Although there have already been some works in this direction
\cite{laszlo}, 
assuming the CDM to be solely due to the BEC, ample scope remains for garnering more
evidence, once the effect of the quantum potential in emulating the DE, alongwith the
backreaction of other constituents (such as baryons), are brought into consideration.

%%%%%%%%%%%%%%%%%%%%%%%%%%%%%%%%%%%%%%%%%%%%%%%%%%%%%%

\bigskip 
\noindent 
{\bf Acknowledgment}

\no
We dedicate this work to the memory of Professor Rajat K. Bhaduri, in collaboration with whom 
the ideas in this paper were first developed. SD acknowledges support of the Natural Sciences 
and Engineering Research Council (NSERC) of Canada. MKS acknowledges financial support of the 
Council of Scientific and Industrial Research (CSIR), Government of India.
%

%%%%%%%%%%%%%%%%%%%%%%%%%%%%%%%%%%%%%%%%%%%%%%%%%%%%%

%\section*{References}

\end{document}